\documentclass[titlepage,11pt]{article}
\usepackage{amsfonts}
\usepackage{amsmath}
\usepackage{amssymb}
\usepackage[onehalfspacing]{setspace}
\usepackage{geometry}
\usepackage{mathrsfs}
\usepackage{boxedminipage}

\newtheorem{theorem}{Theorem}

\input{tcilatex}
\renewcommand{\mathcal}{\mathscr}

\begin{document}

\title{Autocorrelation functions for point-process time series}
\author{Daniel Gervini \\
Department of Mathematical Sciences\\
University of Wisconsin--Milwaukee\\
gervini@uwm.edu}
\maketitle

\begin{abstract}
This article introduces autocorrelograms for time series of point processes.
Such time series usually arise when a longer temporal or spatio-temporal
point process is sliced into smaller time units; for example, when an annual
process is sliced into 365 daily replications. We assume the point processes
follow a doubly-stochastic Poisson model with log-Gaussian intensity
functions. The proposed autocorrelograms are computationally simple and
based on binning. The asymptotic distribution of the autocorrelations is
established. The ability of the method to detect the patterns of common
autoregressive and moving-average time series models is shown by simulation.
Two examples of application to temporal and spatial point-process time
series are shown, pertaining bike demand in the Divvy bike-sharing system of
Chicago and street theft in Chicago, respectively.

\emph{Key Words:} Campbell's theorem; Cox process; functional data analysis;
Poisson process

\emph{MSC classification: 62M10}
\end{abstract}

\section{Introduction}

Spatiotemporal point process models are becoming increasingly common in
statistical applications. The use of point processes for data modeling is
not new (see e.g.~Cox and Isham, 1980; Snyder and Miller, 1991), but the
increasing complexity of today's data and the more powerful computational
resources now available allow for more sophisticated data modelling
approaches than were feasible in the past.

Most of the literature on point-process modelling deals with single
realizations of spatial, temporal or spatiotemporal processes (see
e.g.~Diggle, 2013; M\o ller and Waagepetersen, 2004; Streit, 2010). But in
many applications it is possible to slice a long temporal or spatiotemporal
process, say an annual process, into smaller units, such as daily
occurrences, obtaining in this way repeated observations of a point process.
For example: when the time and location of the occurrences of a certain type
of crime, like street theft, are recorded for a given city on a given year,
this gives rise, in principle, to a single annual spatiotemporal point
process; but this process can be sliced into 365 daily replications of a
spatial process, if the geographical distribution of the incidents is more
important to the researcher than the exact times of the occurrences. These
daily replications, however, are unlikely to be independent; they constitute
a time series of spatial point processes.

The literature on statistical modelling of replicated point processes is
scant, and has only dealt with independent replications (Bouzas et al.,
2006, 2007; Fern\'{a}ndez-Alcal\'{a} et al., 2012; Wu et al., 2013; Gervini,
2016, 2022a, 2022b; Gervini and Khanal, 2019; Gervini and Bauer, 2020). The
first tool that needs to be developed for the analysis of point-process time
series is an analogous of the autocorrelation function, which we propose in
this paper. This is always the first step in time series modelling. Versions
of autocorrelation functions for functional time series have been proposed
recently (Kokoszka and Reimherr, 2013; Mestre et al., 2021; Huang and Shang,
2023), but they cannot be directly applied to point processes, especially
when the realizations are sparse and do not allow estimation of the
intensity functions by smoothing. There are also algebraic operations, like
differencing, which are common for numerical, multivariate or functional
time series but are not allowed for point processes: there is no
statistically meaningful way to take the \textquotedblleft
difference\textquotedblright\ between two sets of points. For these reasons,
the definitions of autocorrelation functions that have been proposed for
functional time series cannot be directly applied to point-process data.

In this paper we have aimed for simplicity and practicality rather than
sophistication, in view of the use of autocorrelograms as basic
data-descriptive tools that point the way to more rigorous data modelling
and statistical inference. The paper is organized as follows: we give a
brief overview of the necessary point process and time series background
(Section \ref{sec:Background}), then introduce autocorrelation functions
based on binning (Section \ref{sec:Binning_estim}), study their ability to
identify various time series models by simulation (Section \ref%
{sec:Simulations}), and finally show two real-data examples of application
(Section \ref{sec:Examples}).

\section{Background on point processes and time series\label{sec:Background}}

A point process $X$ is a random countable set in a space $\mathcal{S}$,
where $\mathcal{S}$ is usually $[0,+\infty )$ for temporal processes or $%
\mathbb{R}^{2}$ for spatial processes (M\o ller and Waagepetersen, 2004,
ch.~2). A point process is locally finite if, for any bounded set $%
B\subseteq \mathcal{S}$, $X\cap B$ is finite with probability one. In that
case, the count function $N(B)=\#(X\cap B)$ is well defined. Given $\lambda
(s)$ a nonnegative locally integrable function, that is, $\lambda :\mathcal{S%
}\rightarrow \lbrack 0,\infty )$ such that $\int_{B}\lambda (s)ds$ is finite
for any bounded $B\subseteq \mathcal{S}$, the process $X$ is said to be a
Poisson process with intensity function $\lambda $, denoted by $X\sim 
\mathcal{P}(\lambda )$, if \emph{(i)} $N(B)$ follows a Poisson distribution
with rate $\int_{B}\lambda (t)dt$ for any bounded $B\subseteq \mathcal{S}$,
and \emph{(ii)} $N(B_{1}),\ldots ,N(B_{k})$ are independent for any
collection of disjoint bounded sets $B_{1},\ldots ,B_{k}$ in $\mathcal{S}$.
A consequence of \emph{(i)} and \emph{(ii)} is that, for a given bounded $%
B\subseteq \mathcal{S}$, the conditional distribution of the points in $%
X\cap B$ given $N(B)=m$ is the distribution of $m$ independent identically
distributed random variables with density $\lambda (s)/\int_{B}\lambda $.

When $n$ Poisson processes $X_{1},\ldots ,X_{n}$ are considered, a single
intensity function $\lambda $ will rarely provide an adequate fit for all of
them; it is more reasonable to assume that each $X_{t}$ is accompanied by a
stochastic process $\Lambda _{t}$ that takes values in the space of
nonnegative locally integrable functions and such that $X_{t}\mid (\Lambda
_{t}=\lambda _{t})\sim \mathcal{P}(\lambda _{t})$. Such pairs $(X,\Lambda )$
are called doubly stochastic or Cox processes (M\o ller and Waagepetersen,
2004, ch.~5). Given $n$ pairs $(X_{1},\Lambda _{1}),\ldots ,(X_{n},\Lambda
_{n})$, the $X_{t}$s are typically observable but the $\Lambda _{t}$s are
not, so they are treated as latent variables in the model. We will assume
throughout this paper that the $X_{t}$s are conditionally independent given
the $\Lambda _{t}$s; therefore, the correlation structure among the $X_{t}$s
is entirely determined by the correlations among the latent $\Lambda _{t}$s.

We assume the $\Lambda _{t}$s are weakly stationary in the index $t$, that
is, that the functions defined by 
\begin{equation*}
\nu \left( s\right) =E\left\{ \Lambda _{t}\left( s\right) \right\} ,
\end{equation*}%
\begin{equation*}
c_{k}\left( s,s^{\prime }\right) =E\left\{ \Lambda _{t}\left( s\right)
\Lambda _{t+k}\left( s^{\prime }\right) \right\} ,\text{ for }k\in \mathbb{Z}%
,
\end{equation*}%
do not depend on $t$. We also assume that the $\Lambda _{t}$s are
log-Gaussian processes: $\Lambda _{t}=\exp \left( G_{t}\right) $ where $%
G_{1},\ldots ,G_{n}$ are weakly stationary Gaussian processes on $\mathcal{S}
$. Let 
\begin{equation*}
\mu \left( s\right) =E\left\{ G_{t}\left( s\right) \right\} ,
\end{equation*}%
\begin{equation*}
\gamma _{k}\left( s,s^{\prime }\right) =\func{cov}\left\{ G_{t}\left(
s\right) ,G_{t+k}\left( s^{\prime }\right) \right\} ,\text{ for }k\in 
\mathbb{Z}.
\end{equation*}%
Expressions for $\nu \left( s\right) $ and $c_{k}\left( s,s^{\prime }\right) 
$ in terms of $\mu \left( s\right) $ and $\gamma _{k}\left( s,s^{\prime
}\right) $ are derived in the Supplementary Material. We have 
\begin{equation}
\nu \left( s\right) =\exp \{\mu \left( s\right) +\frac{1}{2}\gamma
_{0}\left( s,s\right) \},  \label{eq:nu}
\end{equation}%
\begin{equation*}
c_{k}\left( s,s^{\prime }\right) =\nu \left( s\right) \nu \left( s^{\prime
}\right) \exp \{\gamma _{k}\left( s,s^{\prime }\right) \},\text{ for }k\in 
\mathbb{Z}.
\end{equation*}%
Reciprocally, we can express $\mu \left( s\right) $ and $\gamma _{k}\left(
s,s^{\prime }\right) $ in terms of $\nu \left( s\right) $ and $c_{k}\left(
s,s^{\prime }\right) $ as 
\begin{equation}
\mu \left( s\right) =2\log \nu \left( s\right) -\frac{1}{2}\log c_{0}\left(
s,s\right) ,  \label{eq:mu_from_nu}
\end{equation}%
\begin{equation}
\gamma _{k}\left( s,s^{\prime }\right) =\log \left\{ \frac{c_{k}\left(
s,s^{\prime }\right) }{\nu \left( s\right) \nu \left( s^{\prime }\right) }%
\right\} ,\text{ for }k\in \mathbb{Z}.  \label{eq:Gamma_k_from_c}
\end{equation}%
Since the $\Lambda _{t}$s are not observable, $\mu $ and the $\gamma _{k}$s
are not directly estimable; only $\nu $ and the $c_{k}$s can be estimated
directly from the observed $X_{t}$s, and then (\ref{eq:mu_from_nu}) and (\ref%
{eq:Gamma_k_from_c}) can be used to derived estimators for $\mu $ and the $%
\gamma _{k}$s.

By Cauchy-Schwarz's inequality, 
\begin{eqnarray*}
\gamma _{k}^{2}\left( s,s^{\prime }\right) &=&\func{cov}^{2}\left\{
G_{t}\left( s\right) ,G_{t+k}\left( s^{\prime }\right) \right\} \\
&\leq &\func{var}\left\{ G_{t}\left( s\right) \right\} \func{var}\left\{
G_{t+k}\left( s^{\prime }\right) \right\}
\end{eqnarray*}%
for all $k\in \mathbb{Z}$, so, if $v_{0}\left( s\right) $ denotes the
variance function $\gamma _{0}\left( s,s\right) $ of the $G_{t}$s, we have 
\begin{equation*}
\iint_{R\times R}\gamma _{k}^{2}\left( s,s^{\prime }\right) ~ds~ds^{\prime
}\leq \left( \int_{R}v_{0}\left( s\right) ds\right) ^{2}
\end{equation*}%
for any bounded region $R\subset \mathcal{S}$. From now on we are going to
assume that there is a given bounded region $R$ of interest, and that either
all $X_{t}$s take values in $R$ with probability one or that the analysis is
restricted to $X_{t}\cap R$, but for ease of notation we will not indicate
this explicitly. Then, as in Kokoszka and Reimherr (2013) and Mestre et
al.~(2021), the functional autocorrelation coefficient at lag $k$ is defined
as 
\begin{equation}
\tilde{\rho}_{k}=\frac{\left\Vert \gamma _{k}\right\Vert _{2}}{\left\Vert
v_{0}\right\Vert _{1}},  \label{eq:rho_tilde_k}
\end{equation}%
where $\left\Vert \cdot \right\Vert _{2}$ denotes the $L^{2}\left( R\times
R\right) $ norm and $\left\Vert \cdot \right\Vert _{1}$ the $L^{1}\left(
R\right) $ norm. It follows that $0\leq \tilde{\rho}_{k}\leq 1$ for all $k$,
and $\tilde{\rho}_{k}=0$ if and only if $\gamma _{k}\left( s,s^{\prime
}\right) =0$ for all $\left( s,s^{\prime }\right) $. Note that $\gamma
_{k}\left( s,s^{\prime }\right) =\gamma _{-k}\left( s^{\prime },s\right) $
for all $k\in \mathbb{Z}$, so $\left\Vert \gamma _{k}\right\Vert
_{2}=\left\Vert \gamma _{-k}\right\Vert _{2}$ and then $\tilde{\rho}_{k}=%
\tilde{\rho}_{-k}$ for all $k\in \mathbb{Z}$. Therefore we only need to
consider $\tilde{\rho}_{k}$s for $k\geq 0$.

\section{Binned autocorrelations\label{sec:Binning_estim}}

Let us partition the region $R$ into $d$ non-overlapping subregions $%
R_{1},\ldots ,R_{d}$ (subintervals for the temporal case or rectangles for
the spatial case), of equal length or area, and define the respective bin
counts $Y_{tj}=\#\left( X_{t}\cap R_{j}\right) $. Let $\mathbf{Y}_{t}$
denote the vector with elements $Y_{t1},\ldots ,Y_{td}$. Define $\mathbf{%
\hat{\nu}}=\mathbf{\bar{Y}}$, 
\begin{equation*}
\mathbf{\hat{C}}_{0}=\frac{1}{n}\sum_{t=1}^{n}\mathbf{Y}_{t}\mathbf{Y}%
_{t}^{T}-\func{diag}\left( \mathbf{\bar{Y}}\right) ,
\end{equation*}%
where $\func{diag}\left( \mathbf{\bar{Y}}\right) $ denotes the diagonal
matrix with diagonal $\mathbf{\bar{Y}}$, and 
\begin{equation*}
\mathbf{\hat{C}}_{k}=\frac{1}{n-k}\sum_{t=1}^{n-k}\mathbf{Y}_{t}\mathbf{Y}%
_{t+k}^{T},\ \ k\geq 1.
\end{equation*}%
The following results, which are proved in the Supplementary Material,
establish the limiting behavior of these estimators. First we make the
following assumptions:

\begin{description}
\item[A1] $X_{t}\mid (\Lambda _{t}=\lambda _{t})\sim \mathcal{P}(\lambda
_{t})$, the $X_{t}$s are conditionally independent given the $\Lambda _{t}$%
s, and $\left\{ \log \Lambda _{t}\right\} $ is a stationary Gaussian process
with mean $\mu \left( s\right) $ and autocovariances $\gamma _{k}\left(
s,s^{\prime }\right) $.

\item[A2] $\left\Vert \mu \right\Vert _{\infty }<\infty $, $\left\Vert
\gamma _{k}\right\Vert _{\infty }<\infty $ for all $k$, and $%
\lim_{k\rightarrow +\infty }\left\Vert \gamma _{k}\right\Vert _{\infty }=0$
(where $\left\Vert \cdot \right\Vert _{\infty }$ denotes the $\sup $ norm
over the region $R$ or $R\times R$, respectively.)
\end{description}

\begin{theorem}
\label{thm:Consistency}Under assumptions A1 and A2, we have:

\begin{enumerate}
\item $\mathbf{\hat{\nu}}\overset{P}{\longrightarrow }\mathbf{\nu }$ as $%
n\longrightarrow \infty $, where $\mathbf{\nu }=E\left( \mathbf{Y}%
_{t}\right) $ is the vector with elements $\nu _{j}=\int_{R_{j}}\nu \left(
s\right) ds$, and

\item $\mathbf{\hat{C}}_{k}\overset{P}{\longrightarrow }\mathbf{C}_{k}$ as $%
n\longrightarrow \infty $ for $k\geq 0$, where $\mathbf{C}_{k}$ is the $%
d\times d$ matrix with elements 
\begin{equation*}
C_{k,jj^{\prime }}=\iint_{R_{j}\times R_{j^{\prime }}}c_{k}\left(
s,s^{\prime }\right) ~ds~ds^{\prime }.
\end{equation*}
\end{enumerate}
\end{theorem}

Theorem \ref{thm:Consistency} together with equations (\ref%
{eq:Gamma_k_from_c}) and (\ref{eq:rho_tilde_k}) motivate the following
definitions: for $k\geq 0$, let 
\begin{equation*}
\mathbf{\hat{\Gamma}}_{k}:=\log \{\func{diag}(\mathbf{\hat{\nu}}^{-1})%
\mathbf{\hat{C}}_{k}\func{diag}(\mathbf{\hat{\nu}}^{-1})\},
\end{equation*}%
where the vector inverse $\mathbf{\hat{\nu}}^{-1}$ and the logarithm are
understood in a component-wise manner, and define the sample autocorrelation
coefficient as 
\begin{equation}
\hat{\rho}_{k}:=\frac{\Vert \hat{\Gamma}_{k}\Vert _{F}}{\func{tr}\hat{\Gamma}%
_{0}},  \label{eq:rho_k_hat}
\end{equation}%
where $\left\Vert \cdot \right\Vert _{F}$ denotes the Frobenius matrix norm
and $\func{tr}\left( \cdot \right) $ the matrix trace.

\begin{theorem}
\label{thm:Consistency_2}Under assumptions A1 and A2, we have:

\begin{enumerate}
\item $\mathbf{\hat{\Gamma}}_{k}\overset{P}{\longrightarrow }\mathbf{\Gamma }%
_{k}$ as $n\longrightarrow \infty $, where $\mathbf{\Gamma }_{k}$ is the $%
d\times d$ matrix with elements 
\begin{equation*}
\Gamma _{k,jj^{\prime }}=\log \left( \frac{C_{k,jj^{\prime }}}{\nu _{j}\nu
_{j^{\prime }}}\right) ,
\end{equation*}

\item $\hat{\rho}_{k}\overset{P}{\longrightarrow }\rho _{k}$ as $%
n\longrightarrow \infty $, where 
\begin{equation}
\rho _{k}=\frac{\Vert \Gamma _{k}\Vert _{F}}{\func{tr}\Gamma _{0}}.
\label{eq:rho_k}
\end{equation}%
If the autocovariance function $\gamma _{k}$ is zero for any $k\geq 1$, then 
$\mathbf{\Gamma }_{k}=\mathbf{O}$ and $\rho _{k}=0$.
\end{enumerate}
\end{theorem}

Heuristically, the definitions of $\mathbf{\hat{\Gamma}}_{k}$ and $\hat{\rho}%
_{k}$ are motivated by the fact that, when the number of bins $d$ is large
and the lengths or areas of the $R_{j}$s are small, we have $\Gamma
_{k,jj^{\prime }}\approx \gamma _{k}(s,s^{\prime })$ for $(s,s^{\prime })\in
R_{j}\times R_{j^{\prime }}$ and then $\rho _{k}\approx \tilde{\rho}_{k}$.
However, estimation of $\gamma _{k}(s,s^{\prime })$ per se is not the aim of
this paper; our goal is the detection of patterns of decay of the
autocorrelations and the identification of time-series models, such as the
ones considered in Section \ref{sec:Simulations}. The precise magnitude of $%
\left\vert \rho _{k}-\tilde{\rho}_{k}\right\vert $ as a function of $d$ is
not relevant for our purposes, as long as the pattern of decay of the $\rho
_{k}$s resembles that of the $\tilde{\rho}_{k}$s close enough that the right
model can be identified; this is studied by simulation in Section \ref%
{sec:Simulations}.

Broadly speaking, binning allows us to detect autocorrelations in the
variation of the size of the $\Lambda _{t}$s over smaller regions $R_{j}$ of 
$R$. In many cases the autocorrelations emerge in the total counts over $R$
themselves, so that choosing $d=1$ bin is enough to detect them. In other
cases, as in one of the examples of Section \ref{sec:Examples}, the
variation among the $\Lambda _{t}$s consists mainly of a redistribution of
mass over different regions of $R$ while the total mass over $R$ remains
largely unchanged; this kind of variability will be detected only by $d$s
larger than one. However, $d$ need not be large in practice; for real-data
applications, the most relevant types of variability and association among
the $\Lambda _{t}$s will generally emerge for $d$s as low as five, in the
temporal case, or nine, in the spatial case. Using larger numbers of bins
will only increase the variability of the estimators $\hat{\rho}_{k}$ at the
risk of masking their statistical significance.

The next theorem gives the asymptotic distribution of the $\hat{\rho}_{k}$s
when the time series is uncorrelated. This allows us to determine which $%
\hat{\rho}_{k}$s are significantly different from zero. In this theorem $%
\mathbf{I}_{d^{2}}$ denotes the $d^{2}\times d^{2}$ identity matrix, $%
\mathbf{1}_{d}$ the $d$-dimensional vector of ones, and $\otimes $ the
Kronecker product (Magnus and Neudecker, 1999).

\begin{theorem}
\label{thm:Asymp_rhok}Under assumptions A1 and A2, if $\gamma _{h}\equiv 0$
for all $h\neq 0$, then 
\begin{equation*}
n\hat{\rho}_{k}^{2}\overset{D}{\longrightarrow }\frac{\mathbf{Z}^{T}\mathbf{%
B\Omega B}^{T}\mathbf{Z}}{\left( \func{tr}\mathbf{\Gamma }_{0}\right) ^{2}}%
\text{ as }n\longrightarrow \infty ,\ \ \text{for }k\geq 1,
\end{equation*}%
where $\mathbf{Z}\sim N\left( \mathbf{0},\mathbf{I}_{d^{2}}\right) $, $%
\mathbf{B}=\left[ \mathbf{B}_{1},\mathbf{B}_{2}\right] $ is the $d^{2}\times
\left( d+d^{2}\right) $ matrix with blocks 
\begin{eqnarray*}
\mathbf{B}_{1} &=&-\left( \mathbf{1}_{d}\otimes \func{diag}\left( \mathbf{%
\nu }^{-1}\right) \right) -\left( \func{diag}\left( \mathbf{\nu }%
^{-1}\right) \otimes \mathbf{1}_{d}\right) , \\
\mathbf{B}_{2} &=&\func{diag}\left( \mathbf{\nu }^{-1}\otimes \mathbf{\nu }%
^{-1}\right) ,
\end{eqnarray*}%
and $\mathbf{\Omega }$ is the $\left( d+d^{2}\right) \times \left(
d+d^{2}\right) $ matrix with block structure 
\begin{equation*}
\mathbf{\Omega }=\left[ 
\begin{array}{cc}
\mathbf{\Omega }_{11} & \mathbf{\Omega }_{12} \\ 
\mathbf{\Omega }_{12}^{T} & \mathbf{\Omega }_{22}%
\end{array}%
\right]
\end{equation*}%
where $\mathbf{\Omega }_{11}=E\{\left( \mathbf{Y}_{t}-\mathbf{\nu }\right)
\left( \mathbf{Y}_{t}-\mathbf{\nu }\right) ^{T}\}$, $\mathbf{\Omega }_{12}=%
\mathbf{\nu }^{T}\otimes \mathbf{\Omega }_{11}+\mathbf{\Omega }_{11}\otimes 
\mathbf{\nu }^{T}$ and 
\begin{eqnarray*}
\mathbf{\Omega }_{22} &=&\mathbf{\Omega }_{11}\otimes \mathbf{\Omega }_{11}+%
\mathbf{\Omega }_{11}\otimes \mathbf{\nu \nu }^{T}+\mathbf{\nu \nu }%
^{T}\otimes \mathbf{\Omega }_{11}+ \\
&&\left( \mathbf{\nu }\otimes \mathbf{I}_{d}\right) \mathbf{\Omega }%
_{11}\left( \mathbf{I}_{d}\otimes \mathbf{\nu }^{T}\right) +\left( \mathbf{I}%
_{d}\otimes \mathbf{\nu }\right) \mathbf{\Omega }_{11}\left( \mathbf{\nu }%
^{T}\otimes \mathbf{I}_{d}\right) .
\end{eqnarray*}
\end{theorem}

From Theorem \ref{thm:Asymp_rhok} we can derive asymptotic upper confidence
bounds for $\hat{\rho}_{k}$ under the hypothesis of independence. The
parameters $\mathbf{\nu }$ and $\mathbf{\Omega }_{11}$ are estimated by the
sample mean and the sample covariance matrix of the $\mathbf{Y}_{t}$s,
respectively, and the matrix $\mathbf{\Gamma }_{0}$ by $\mathbf{\hat{\Gamma}}%
_{0}$. Then the $1-\alpha $ quantile of the asymptotic distribution of $n%
\hat{\rho}_{k}^{2}$ can be obtained by Monte Carlo.

In the next section we investigate, by simulation, the ability of the $\hat{%
\rho}_{k}$s to detect the typical patterns of some of the most common time
series models, as well as the finite-sample behavior of the asymptotic
confidence bounds.

\section{Simulations\label{sec:Simulations}}

\subsection{Finite-dimensional processes}

For most applications it can be assumed that the $G_{t}$s belong to a
finite-dimensional functional space. Then we can write, without loss of
generality, 
\begin{equation}
G_{t}\left( s\right) =\mu \left( s\right) +\mathbf{U}_{t}^{T}\mathbf{\phi }%
\left( s\right) ,  \label{eq:Finite_dim_model}
\end{equation}%
where $\mathbf{\phi }\left( s\right) $ is a vector of $p$ orthogonal
functions in $L^{2}\left( R\right) $ and $\{\mathbf{U}_{t}\}$ is a
stationary zero-mean $p$-variate time series. In that case 
\begin{eqnarray*}
\gamma _{k}\left( s,s^{\prime }\right) &=&E\left\{ \mathbf{U}_{t}^{T}\mathbf{%
\phi }\left( s\right) \mathbf{U}_{t+k}^{T}\mathbf{\phi }\left( s^{\prime
}\right) \right\} \\
&=&\mathbf{\phi }\left( s\right) ^{T}\mathbf{\Sigma }_{k}\mathbf{\phi }%
\left( s^{\prime }\right) ,
\end{eqnarray*}%
where $\mathbf{\Sigma }_{k}=E\left( \mathbf{U}_{t}\mathbf{U}%
_{t+k}^{T}\right) $. Note that $\mathbf{\Sigma }_{-k}=\mathbf{\Sigma }%
_{k}^{T}$. Then, after some algebra, we get $\iint \gamma _{k}^{2}\left(
s,s^{\prime }\right) ~ds~ds^{\prime }=\left\Vert \mathbf{\Sigma }%
_{k}\right\Vert _{F}^{2}$ and $\int v_{0}\left( s\right) ds=\func{tr}\mathbf{%
\Sigma }_{0}$, so $\tilde{\rho}_{k}=\left\Vert \mathbf{\Sigma }%
_{k}\right\Vert _{F}/\func{tr}\mathbf{\Sigma }_{0}$ in this case.

Two simple models of practical interest are the moving average and the
autoregressive models of order one, denoted by $\mathrm{MA}\left( 1\right) $
and $\mathrm{AR}\left( 1\right) $ respectively. The $\mathrm{MA}\left(
1\right) $ model is given by $\mathbf{U}_{t}=\mathbf{Z}_{t}+\mathbf{BZ}%
_{t-1} $, where $\{\mathbf{Z}_{t}\}$ are zero-mean uncorrelated (across
different $t $s) random vectors with $\mathbf{V}=E\left( \mathbf{Z}_{t}%
\mathbf{Z}_{t}^{T}\right) $. For this model we have $\mathbf{\Sigma }_{0}=%
\mathbf{V}+\mathbf{BVB}^{T}$, $\mathbf{\Sigma }_{1}=\mathbf{VB}^{T}$ and $%
\mathbf{\Sigma }_{k}=\mathbf{O}$ for $k\geq 2$. Then 
\begin{equation*}
\tilde{\rho}_{1}=\frac{\left\Vert \mathbf{VB}^{T}\right\Vert _{F}}{\func{tr}%
\left( \mathbf{V}+\mathbf{BVB}^{T}\right) }\text{ and }\tilde{\rho}_{k}=0%
\text{ for }k\geq 2.
\end{equation*}

The $\mathrm{AR}\left( 1\right) $ model is given by $\mathbf{U}_{t}=\mathbf{%
AU}_{t-1}+\mathbf{Z}_{t}$, where the $\mathbf{Z}_{t}$s are as before. If $%
\left\Vert \mathbf{A}\right\Vert _{F}<1$, then $\mathbf{U}%
_{t}=\sum_{h=0}^{\infty }\mathbf{A}^{h}\mathbf{Z}_{t-h}$, $\mathbf{\Sigma }%
_{0}=\sum_{h=0}^{\infty }\mathbf{A}^{h}\mathbf{V}(\mathbf{A}^{h})^{T}$ and $%
\mathbf{\Sigma }_{k}=\sum_{h=0}^{\infty }\mathbf{A}^{h}\mathbf{V(A}%
^{k+h})^{T}=\mathbf{\Sigma }_{0}\mathbf{(A}^{k})^{T}$ for $k\geq 1$, so 
\begin{equation*}
\tilde{\rho}_{k}=\frac{\left\Vert \mathbf{A}^{k}\mathbf{\Sigma }%
_{0}\right\Vert _{F}}{\func{tr}\mathbf{\Sigma }_{0}}\text{ for }k\geq 1.
\end{equation*}%
Since $\left\Vert \mathbf{A}^{k}\mathbf{\Sigma }_{0}\right\Vert _{F}\leq
\left\Vert \mathbf{A}\right\Vert _{F}^{k}\left\Vert \mathbf{\Sigma }%
_{0}\right\Vert _{F}$ and $\func{tr}\mathbf{\Sigma }_{0}\geq \left\Vert 
\mathbf{\Sigma }_{0}\right\Vert _{F}$, it follows that $\tilde{\rho}_{k}$
decreases exponentially as $k$ increases.

In addition to these models, also common in practice are their respective
seasonal versions denoted by $\mathrm{SMA}_{\tau }\left( 1\right) $ and $%
\mathrm{SAR}_{\tau }\left( 1\right) $, where $\tau $ is the seasonal period.
The $\mathrm{SMA}_{\tau }\left( 1\right) $ model is given by $\mathbf{U}_{t}=%
\mathbf{Z}_{t}+\mathbf{BZ}_{t-\tau }$, where the $\mathbf{Z}_{t}$s are as
before. For this model we have $\mathbf{\Sigma }_{0}=\mathbf{V}+\mathbf{BVB}%
^{T}$, $\mathbf{\Sigma }_{\tau }=\mathbf{VB}^{T}$ and $\mathbf{\Sigma }_{k}=%
\mathbf{O}$ for any $k>0$ with $k\neq \tau $.Then 
\begin{equation*}
\tilde{\rho}_{\tau }=\frac{\left\Vert \mathbf{VB}^{T}\right\Vert _{F}}{\func{%
tr}\left( \mathbf{V}+\mathbf{BVB}^{T}\right) }\text{ and }\tilde{\rho}_{k}=0%
\text{ for any }k\neq \tau .
\end{equation*}

The $\mathrm{SAR}_{\tau }\left( 1\right) $ model is given by $\mathbf{U}_{t}=%
\mathbf{AU}_{t-\tau }+\mathbf{Z}_{t}$, where the $\mathbf{Z}_{t}$s are as
before. If $\left\Vert \mathbf{A}\right\Vert _{F}<1$ then $\mathbf{U}%
_{t}=\sum_{h=0}^{\infty }\mathbf{A}^{h}\mathbf{Z}_{t-\tau h}$, $\mathbf{%
\Sigma }_{0}=\sum_{h=0}^{\infty }\mathbf{A}^{h}\mathbf{V}(\mathbf{A}%
^{h})^{T} $ and, for $k$ a multiple of $\tau $, $\mathbf{\Sigma }_{k}=%
\mathbf{\Sigma }_{0}(\mathbf{A}^{k/\tau })^{T}$. All other $\mathbf{\Sigma }%
_{k}$s, for $k$ not a multiple of $\tau $, are zero. Then 
\begin{equation*}
\tilde{\rho}_{k\tau }=\frac{\left\Vert \mathbf{A}^{k}\mathbf{\Sigma }%
_{0}\right\Vert _{F}}{\func{tr}\mathbf{\Sigma }_{0}}\text{ and }\tilde{\rho}%
_{h}=0\text{ for }h\text{ not a multiple of }\tau .
\end{equation*}%
As before, $\tilde{\rho}_{k\tau }$ decreases exponentially as $k$ increases.

The simulations that follow will focus on these four models.

\subsection{Number of bins, time series length and pattern detection}

The first set of simulations was designed to investigate the relationship
between estimation error of the $\hat{\rho}_{k}$s, the number of bins $d$
and the time series length $n$. We know by Theorem \ref{thm:Consistency_2}
that, for fixed $d$, the $\hat{\rho}_{k}$s are not consistent estimators of
the autocorrelation coefficients $\tilde{\rho}_{k}$ given by (\ref%
{eq:rho_tilde_k}) but of the $\rho _{k}$s given by (\ref{eq:rho_k}).
However, we would expect that, if the bias is not too large, the $\hat{\rho}%
_{k}$s would still be useful to detect the typical patterns of decay of the $%
\tilde{\rho}_{k}$s for the most common time series models.

We considered different $\mathrm{MA}\left( 1\right) $, $\mathrm{AR}(1)$, $%
\mathrm{SMA}_{\tau }\left( 1\right) $ and $\mathrm{SAR}_{\tau }(1)$ models.
For simplicity we only simulated one-dimensional models, that is, models (%
\ref{eq:Finite_dim_model}) with $p=1$. Then, for $\mathrm{MA}\left( 1\right) 
$\ models we have $\sigma _{0}=v+vb^{2}$ and $\sigma _{1}=vb$, so 
\begin{equation*}
\tilde{\rho}_{1}=\frac{\left\vert vb\right\vert }{v+vb^{2}}=\frac{\left\vert
b\right\vert }{1+b^{2}}.
\end{equation*}%
This $\tilde{\rho}_{1}$ never exceeds $0.5$. We chose two values of $b$ for
the simulations, namely $b\in \left\{ 1,3.732\right\} $, which give $\tilde{%
\rho}_{1}\in \left\{ 0.5,0.25\right\} $ respectively. For $\mathrm{AR}\left(
1\right) $ models we have 
\begin{equation*}
\sigma _{0}=v\sum_{h=0}^{\infty }a^{2k}=\frac{v}{1-a^{2}}
\end{equation*}%
and 
\begin{equation*}
\tilde{\rho}_{k}=\frac{\sigma _{0}\left\vert a\right\vert ^{k}}{\sigma _{0}}%
=\left\vert a\right\vert ^{k}.
\end{equation*}%
We chose four different values of $a$, namely $a\in \left\{
0,0.25,0.50,0.75\right\} $.

The white noise variance $v$ for the above models, as well as the mean
function $\mu \left( s\right) $ and the variance component $\phi \left(
s\right) $, were chosen so that the total expected count, 
\begin{equation*}
\int \nu \left( s\right) ds=\int \exp \{\mu \left( s\right) +\frac{1}{2}%
\sigma _{0}\phi \left( s\right) ^{2}\}ds,
\end{equation*}%
was a moderate number, and the largest count likely to occur, 
\begin{equation*}
\int \exp \left\{ \mu \left( s\right) +2\sqrt{\sigma _{0}}\phi \left(
s\right) \right\} ds,
\end{equation*}%
was not too large. Then we took $\mu \left( s\right) =3$ and $\phi \left(
s\right) =\sqrt{2}\sin \left( 2\pi s\right) $ for $s\in \left[ 0,1\right] $,
which, for $\sigma _{0}=1$, give a total expected count of $35.2$ and a
largest count likely to occur of $85.4$, which are reasonable numbers given
what we see in applications (such as those mentioned in Section \ref%
{sec:Examples}). The white-noise variance $v$ was then chosen so as to keep
the total variance $\sigma _{0}$ constant at $1$; therefore, for the $%
\mathrm{MA}\left( 1\right) $\ models we took $v=1/\left( 1+b^{2}\right) $
and for the $\mathrm{AR}\left( 1\right) $\ models we took $v=1-a^{2}$.

For the seasonal models $\mathrm{SMA}_{\tau }\left( 1\right) $ and $\mathrm{%
SAR}_{\tau }(1)$ we chose a seasonal period $\tau =5$, which does not
require a very large sample size $n$ to be detected. The other parameters
were the same as for the respective $\mathrm{MA}\left( 1\right) $ and $%
\mathrm{AR}\left( 1\right) $\ models.

As sample sizes we took $n\in \left\{ 50,100,150,\ldots ,400\right\} $ for $%
\mathrm{MA}\left( 1\right) $ and $\mathrm{AR}\left( 1\right) $\ models, and $%
n\in \left\{ 100,150,\ldots ,400\right\} $ for $\mathrm{SMA}_{5}\left(
1\right) $ and $\mathrm{SAR}_{5}\left( 1\right) $\ models. Each model was
replicated $500$ times. Tables with mean absolute errors $E\left( \left\vert 
\hat{\rho}_{1}-\tilde{\rho}_{1}\right\vert \right) $ and $%
E(\sum_{k=1}^{5}\left\vert \hat{\rho}_{k}-\tilde{\rho}_{k}\right\vert /5)$
are given in the Supplementary Material. Here we only show the expected
autocorrelograms, $E(\hat{\rho}_{k})$ as a function of $k$, in Figures \ref%
{fig:plots_ar}--\ref{fig:plots_sma}.

\FRAME{ftbpFU}{5.7372in}{4.1174in}{0pt}{\Qcb{Simulation Results. Expected
autocorrelograms for $\mathrm{AR}(1)$ models with parameters (a) $a=0$, (b) $%
a=0.25$, (c) $a=0.50$ and (d) $a=0.75$, for time series lengths $n=50$
(circles), $n=100$ (triangles), $n=200$ (squares) and $n=400$ (diamonds).
True autocorrelations $\tilde{\protect\rho}_{k}$ are shown as solid circles.}%
}{\Qlb{fig:plots_ar}}{plots_ar_paper.eps}{\special{language "Scientific
Word";type "GRAPHIC";maintain-aspect-ratio TRUE;display "ICON";valid_file
"F";width 5.7372in;height 4.1174in;depth 0pt;original-width
5.7095in;original-height 4.0906in;cropleft "0";croptop "1";cropright
"1";cropbottom "0";filename '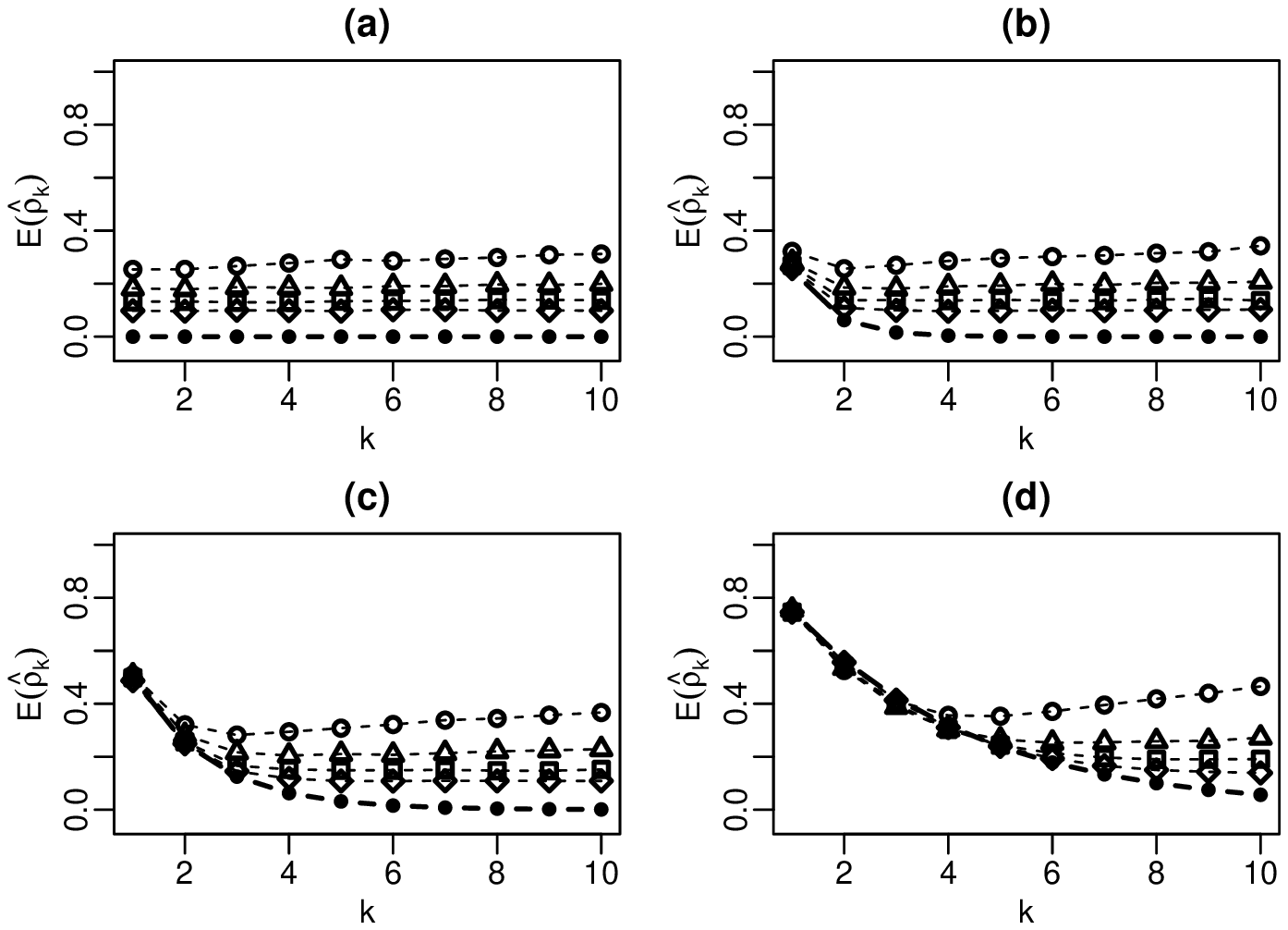';file-properties "XNPEU";}}

\FRAME{ftbpFU}{4.5956in}{3.3001in}{0pt}{\Qcb{Simulation Results. Expected
autocorrelograms for $\mathrm{MA}(1)$ models with parameters (a) $b=1$ and
(b) $b=3.732$, for time series lengths $n=50$ (circles), $n=100$
(triangles), $n=200$ (squares) and $n=400$ (diamonds). True autocorrelations 
$\tilde{\protect\rho}_{k}$ are shown as solid circles.}}{\Qlb{fig:plots_ma}}{%
plots_ma_paper.eps}{\special{language "Scientific Word";type
"GRAPHIC";maintain-aspect-ratio TRUE;display "ICON";valid_file "F";width
4.5956in;height 3.3001in;depth 0pt;original-width 5.7095in;original-height
4.0906in;cropleft "0";croptop "1";cropright "1";cropbottom "0";filename
'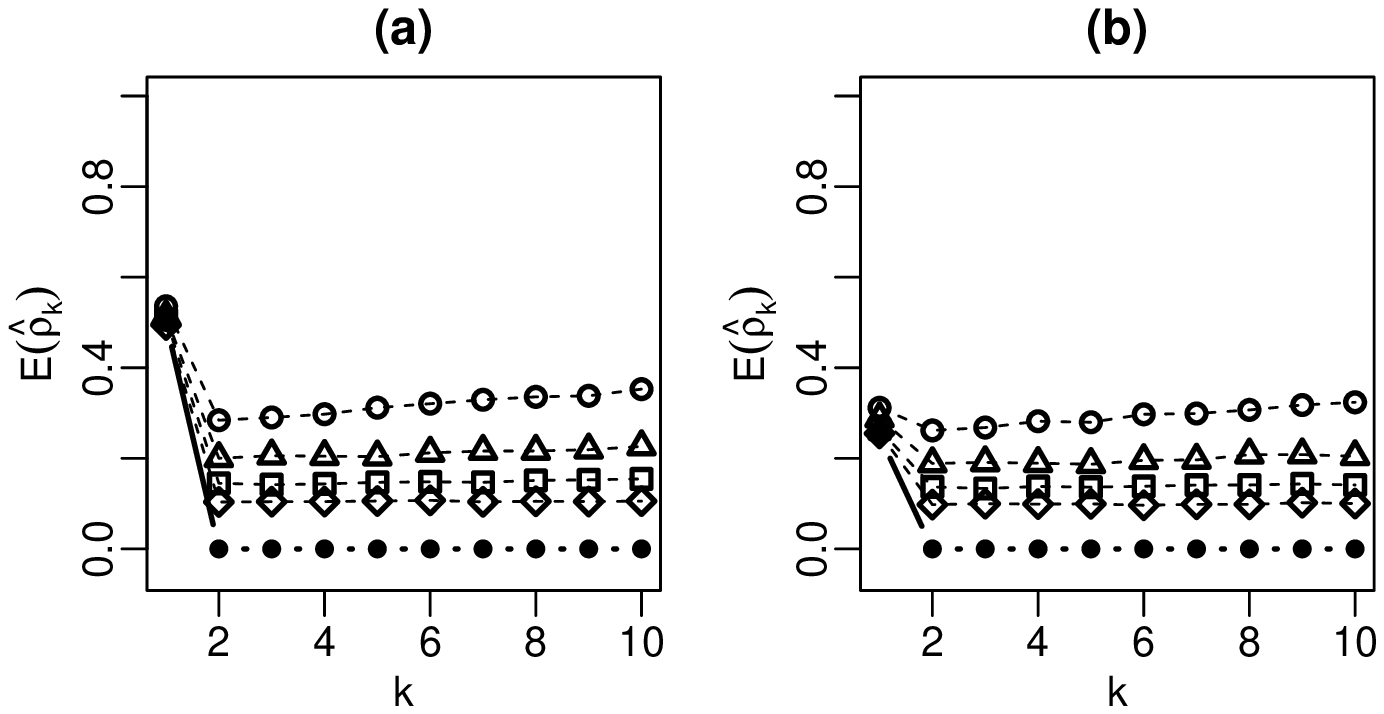';file-properties "XNPEU";}}

\FRAME{ftbpFU}{5.7372in}{4.1174in}{0pt}{\Qcb{Simulation Results. Expected
autocorrelograms for $\mathrm{SAR}_{5}(1)$ models with parameters (a) $%
a=0.25 $, (b) $a=0.50$ and (c) $a=0.75$, for time series lengths $n=100$
(triangles), $n=200$ (squares) and $n=400$ (diamonds). True autocorrelations 
$\tilde{\protect\rho}_{k}$ are shown as solid circles.}}{\Qlb{fig:plots_sar}%
}{plots_sar_paper.eps}{\special{language "Scientific Word";type
"GRAPHIC";maintain-aspect-ratio TRUE;display "ICON";valid_file "F";width
5.7372in;height 4.1174in;depth 0pt;original-width 5.7095in;original-height
4.0906in;cropleft "0";croptop "1";cropright "1";cropbottom "0";filename
'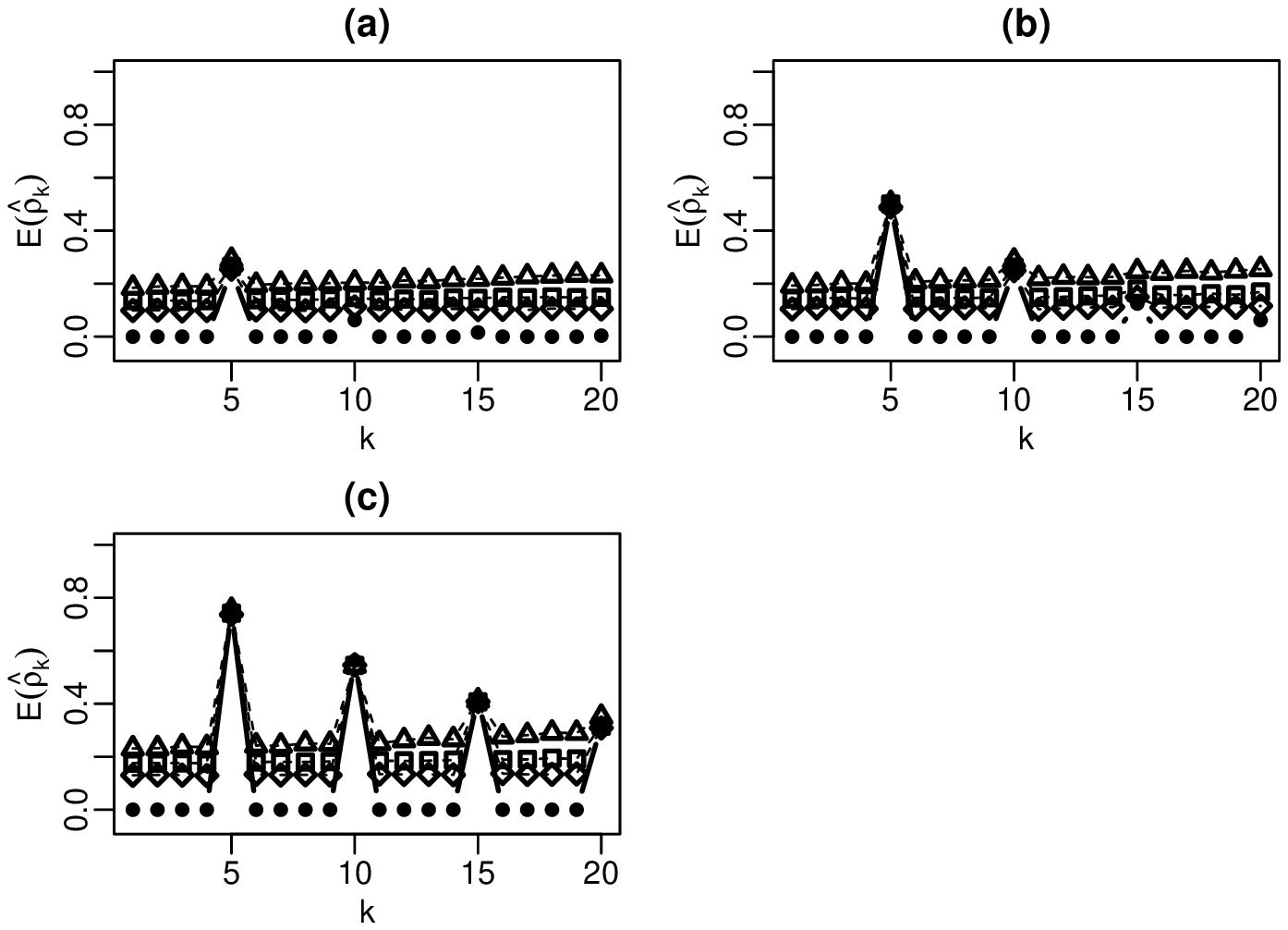';file-properties "XNPEU";}}

\FRAME{ftbpFU}{4.5679in}{3.2724in}{0pt}{\Qcb{Simulation Results. Expected
autocorrelograms for $\mathrm{SMA}_{5}(1)$ models with parameters (a) $b=1$
and (b) $b=3.732$, for time series lengths $n=100$ (triangles), $n=200$
(squares) and $n=400$ (diamonds). True autocorrelations $\tilde{\protect\rho}%
_{k}$ are shown as solid circles.}}{\Qlb{fig:plots_sma}}{plots_sma_paper.eps%
}{\special{language "Scientific Word";type "GRAPHIC";maintain-aspect-ratio
TRUE;display "ICON";valid_file "F";width 4.5679in;height 3.2724in;depth
0pt;original-width 5.7095in;original-height 4.0906in;cropleft "0";croptop
"1";cropright "1";cropbottom "0";filename
'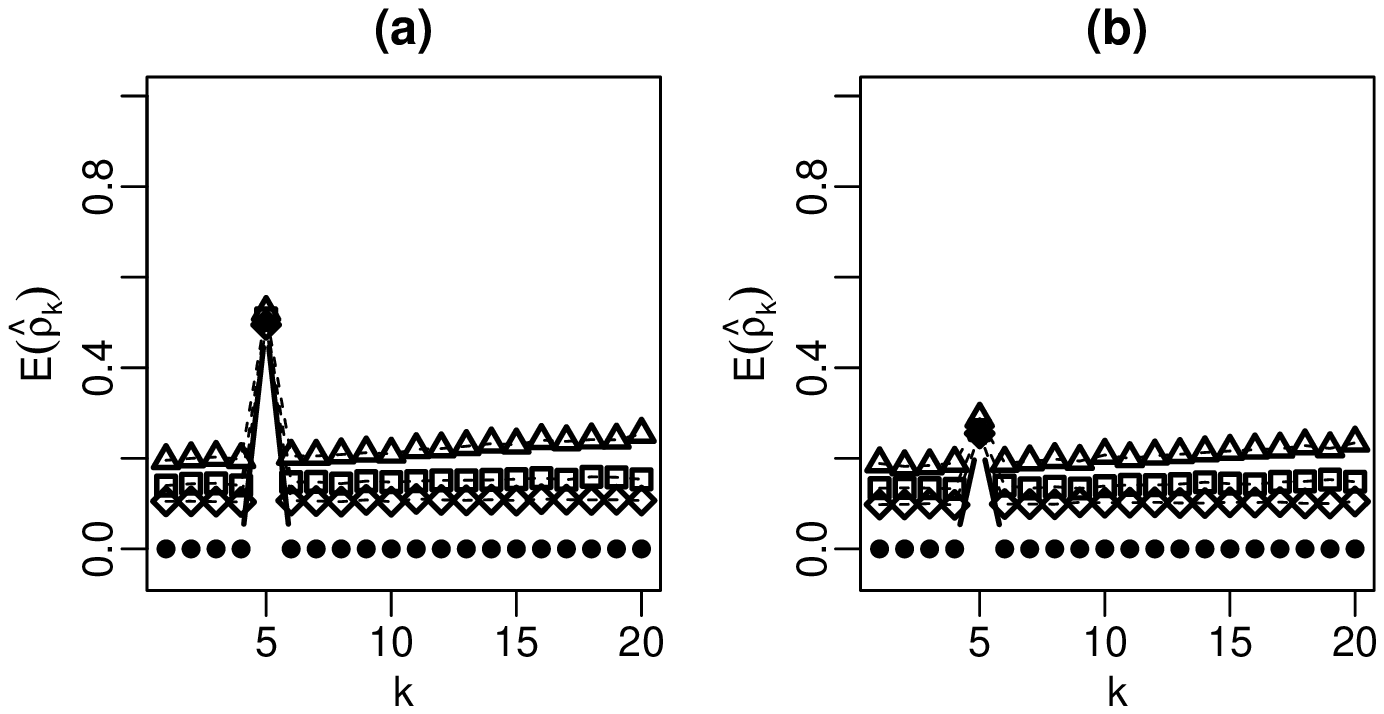';file-properties "XNPEU";}}

From the tables in the Supplementary Material we see that the estimators
based on the smallest number of bins, $d=5$, have the lowest mean absolute
errors in all cases. Figures \ref{fig:plots_ar}--\ref{fig:plots_sma}, then,
show expected autocorrelograms for estimators based on $d=5$ bins. In Figure %
\ref{fig:plots_ar} the pattern of decay of an $\mathrm{AR}(1)$ model is
clearly recognizable, especially for $n\geq 100$. Naturally, it is more
easily recognizable for models with higher correlation parameters $a$. The
pattern of a $\mathrm{MA}\left( 1\right) $ model is also clearly detected in
Figure \ref{fig:plots_ma}, especially for $n\geq 100$. As for the seasonal
models, it is clear from Figures \ref{fig:plots_sar} and \ref{fig:plots_sma}
that an isolated peak is present at lag 5, which immediately suggests
seasonality. Detecting the difference between a $\mathrm{SMA}_{5}\left(
1\right) $ and a $\mathrm{SAR}_{5}(1)$ pattern, however, is harder, but the
distinctive $\mathrm{SAR}_{5}(1)$ pattern is evident in Figure \ref%
{fig:plots_sar} for models with $a=0.75$ if $n\geq 100$, or $a\geq 0.50$ if $%
n\geq 200$.

As mentioned earlier, estimation of $\tilde{\rho}_{k}$ per se is not the
main goal of this paper, but we did look more closely into the convergence
rates of the bias and standard deviation of the $\hat{\rho}_{k}$s when both $%
d$ and $n$ increase, at least for some models. Specifically, for the $%
\mathrm{AR}(1)$ model with $a=0.75$ and the $\mathrm{MA}\left( 1\right) $
model with $b=1$, we generated data with $n\in \left\{
100,200,300,400,500\right\} $ and computed estimators with bin sizes $d\in
\left\{ 2,4,6,\ldots ,16\right\} $. Each scenario was replicated $1000$
times. Tables with biases and standard deviations for $\hat{\rho}_{1}$ are
given in the Supplementary Material. To summarize the results here, we found
out that if we assume $\left\vert \mathrm{bias}\left( \hat{\rho}_{1}\right)
\right\vert =cn^{\alpha }d^{\beta }$, then the coefficients that best fit
our simulations are $\alpha =-0.38$ and $\beta =0.03$ for the $\mathrm{AR}%
(1) $ model and $\alpha =-0.46$ and $d=0.22$ for the $\mathrm{MA}\left(
1\right) $ model. Similarly, if we assume $\mathrm{std}\left( \hat{\rho}%
_{1}\right) =cn^{\alpha }d^{\beta }$, then we get $\alpha =-0.48$ and $\beta
=0.08$ for the $\mathrm{AR}(1)$ model and $\alpha =-0.36$ and $\beta =0.02$
for the $\mathrm{MA}\left( 1\right) $ model. These rates, as we can see,
depend very little on $d$ and are close to the parametric convergence rate
of $\alpha =-0.5$, at least within the range of $d$s and $n$s considered.
Nevertheless, if estimation of the autocovariance functions is desired,
better methods based on smoothing can be developed; but that goes beyond the
scope of this paper.

\subsection{Significance testing}

We also studied by simulation the finite-sample behavior of the asymptotic
confidence bounds derived from Theorem \ref{thm:Asymp_rhok}. For the same $%
\mathrm{AR}(1)$, $\mathrm{MA}(1)$, $\mathrm{SAR}_{5}(1)$ and $\mathrm{SMA}%
_{5}(1)$ models described above, we computed 90\% upper confidence bounds, $%
\mathrm{ub}_{.10}$, and estimated the power functions $P\left( \hat{\rho}%
_{k}\geq \mathrm{ub}_{.10}\right) $ by Monte Carlo. The results reported in
Figures \ref{fig:plots_ucb_ar}--\ref{fig:plots_ucb_sma} are for estimators
based on $d=5$ bins.

\FRAME{ftbpFU}{5.9179in}{4.1174in}{0pt}{\Qcb{Simulation Results. Probability
of estimated autocorrelations $\hat{\protect\rho}_{k}$ exceeding the nominal
90\% upper confidence bound for $\mathrm{AR}(1)$ models with parameters (a) $%
a=0$, (b) $a=0.25$, (c) $a=0.50$ and (d) $a=0.75$, for time series lengths $%
n=50$ (circles), $n=100$ (triangles), $n=200$ (squares) and $n=400$
(diamonds).}}{\Qlb{fig:plots_ucb_ar}}{plots_ucb_ar_paper.eps}{\special%
{language "Scientific Word";type "GRAPHIC";maintain-aspect-ratio
TRUE;display "ICON";valid_file "F";width 5.9179in;height 4.1174in;depth
0pt;original-width 5.8894in;original-height 4.0906in;cropleft "0";croptop
"1";cropright "1";cropbottom "0";filename
'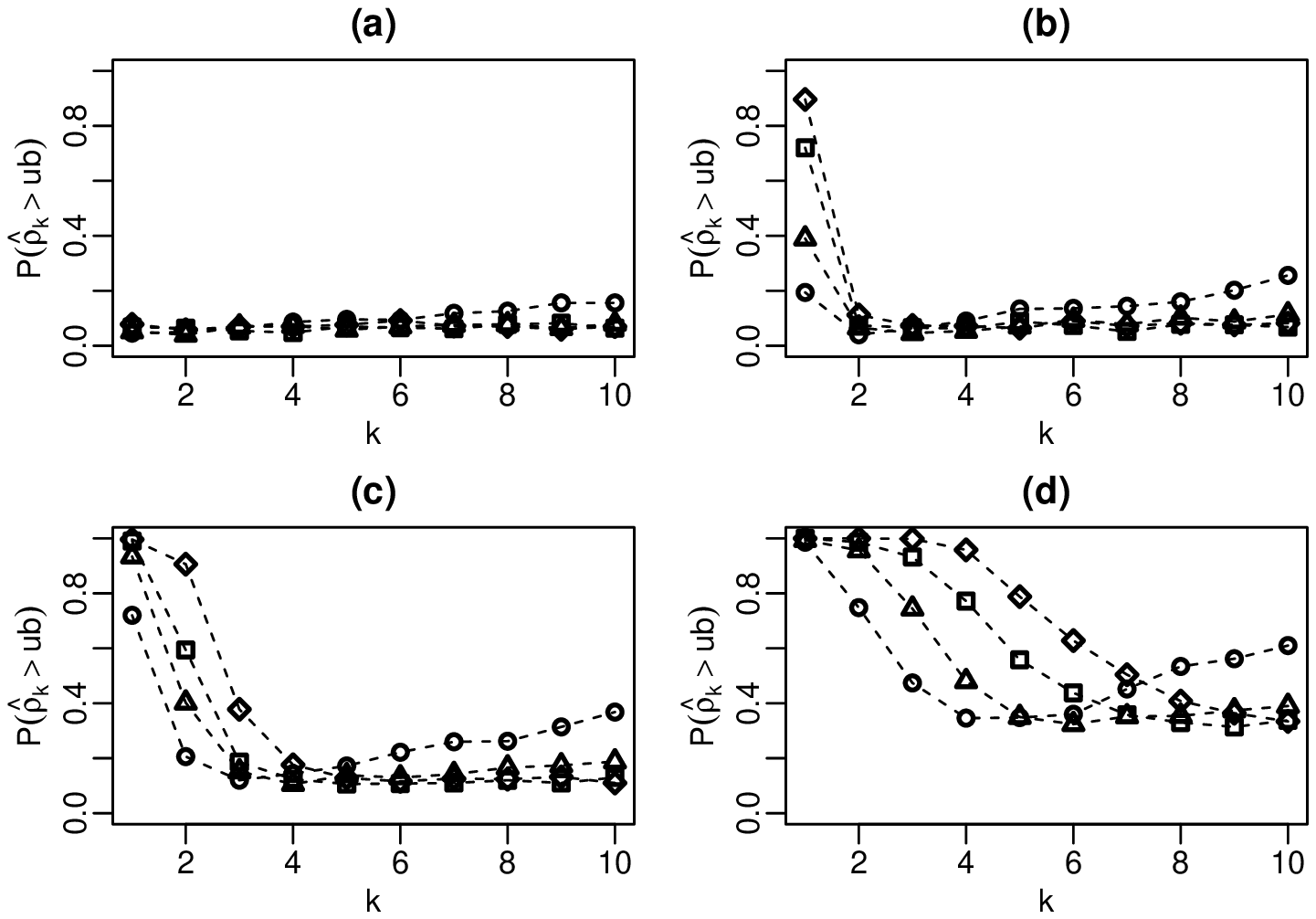';file-properties "XNPEU";}}

\FRAME{ftbpFU}{4.7115in}{3.2716in}{0pt}{\Qcb{Simulation Results. Probability
of estimated autocorrelations $\hat{\protect\rho}_{k}$ exceeding the nominal
90\% upper confidence bound for $\mathrm{MA}(1)$ models with parameters (a) $%
b=1$ and (b) $b=3.732$, for time series lengths $n=50$ (circles), $n=100$
(triangles), $n=200$ (squares) and $n=400$ (diamonds).}}{\Qlb{%
fig:plots_ucb_ma}}{plots_ucb_ma_paper.eps}{\special{language "Scientific
Word";type "GRAPHIC";maintain-aspect-ratio TRUE;display "ICON";valid_file
"F";width 4.7115in;height 3.2716in;depth 0pt;original-width
5.8894in;original-height 4.0906in;cropleft "0";croptop "1";cropright
"1";cropbottom "0";filename '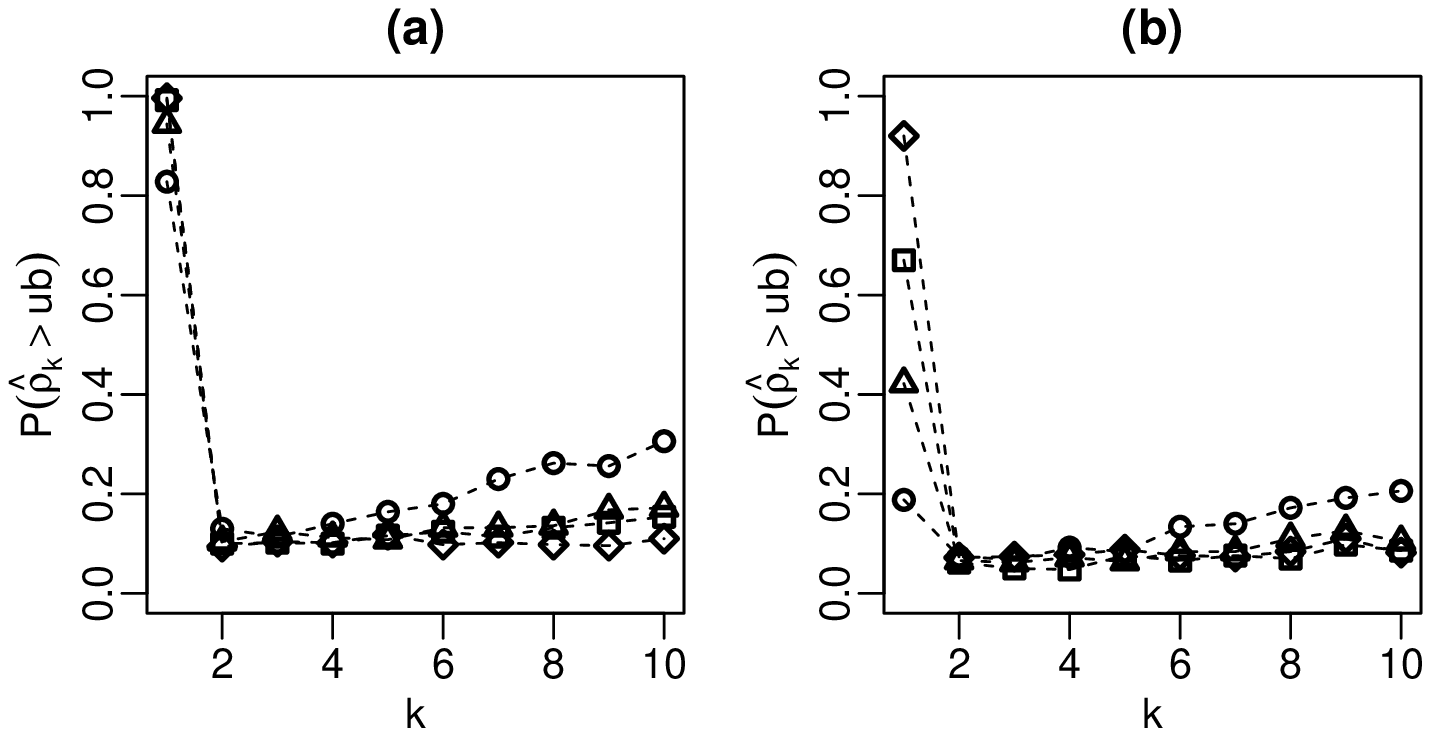';file-properties
"XNPEU";}}

For the white-noise model (Fig.~\ref{fig:plots_ucb_ar}(a)) we see that the
probability of exceeding the upper bound hovers around the nominal 10\%, as
expected; for the case $n=50$ the probabilities become unstable at higher
lags, but this was also expected. For the $\mathrm{AR}(1)$ models with $a>0$%
, the power for detecting the first significant lag depends clearly on $a$
and $n$. For $a=0.25$ (Fig.~\ref{fig:plots_ucb_ar}(b)) the first
autocorrelation $\hat{\rho}_{1}$ will be detected as significant 40\% of the
time if $n=100$ and at least 75\% of the time for $n\geq 200$. On the other
hand, for the moderately correlated model with $a=0.50$ (Fig.~\ref%
{fig:plots_ucb_ar}(c)) the first autocorrelation will be detected as
significant almost 100\% of the time and the second autocorrelation at least
40\% of the time for $n\geq 100$. For the highly correlated model with $%
a=0.75$ (Fig.~\ref{fig:plots_ucb_ar}(d)) the first two autocorrelations will
be detected as significant almost 100\% of the time and the third one at
least 75\% of the time for $n\geq 100$. Therefore, for $\mathrm{AR}(1)$
models with $a\geq 0.50$ and time series lengths $n\geq 100$, the typical $%
\mathrm{AR}(1)$ pattern will be detected with a high probability. Similar
conclusions can be drawn for the seasonal $\mathrm{SAR}_{5}(1)$ models (Fig.~%
\ref{fig:plots_ucb_sar}).

For $\mathrm{MA}(1)$ models, the probability of detecting $\hat{\rho}_{1}$
as significant depends, again, on the strength of the autocorrelation and
the sample size. For the model with $b=1$ (Fig.~\ref{fig:plots_ucb_ma}(a))
for which $\tilde{\rho}_{1}=0.5$, the first autocorrelation $\hat{\rho}_{1}$
is detected as significant about 80\% of the time for $n=50$ and practically
100\% of the time for $n\geq 100$, and all $\hat{\rho}_{k}$s at higher lags
remain at the nominal 10\% level, so the distinct $\mathrm{MA}(1)$ pattern
will be detected without difficulty in this case. Similar conclusions can be
drawn for the seasonal $\mathrm{SMA}_{5}(1)$ models (Fig.~\ref%
{fig:plots_ucb_sma}).

\FRAME{ftbpFU}{5.9179in}{4.1174in}{0pt}{\Qcb{Simulation Results. Probability
of estimated autocorrelations $\hat{\protect\rho}_{k}$ exceeding the nominal
90\% upper confidence bound for $\mathrm{SAR}_{5}(1)$ models with parameters
(a) $a=0.25$, (b) $a=0.50$ and (c) $a=0.75$, for time series lengths $n=100$
(triangles), $n=200$ (squares) and $n=400$ (diamonds).}}{\Qlb{%
fig:plots_ucb_sar}}{plots_ucb_sar_paper.eps}{\special{language "Scientific
Word";type "GRAPHIC";maintain-aspect-ratio TRUE;display "ICON";valid_file
"F";width 5.9179in;height 4.1174in;depth 0pt;original-width
5.8894in;original-height 4.0906in;cropleft "0";croptop "1";cropright
"1";cropbottom "0";filename '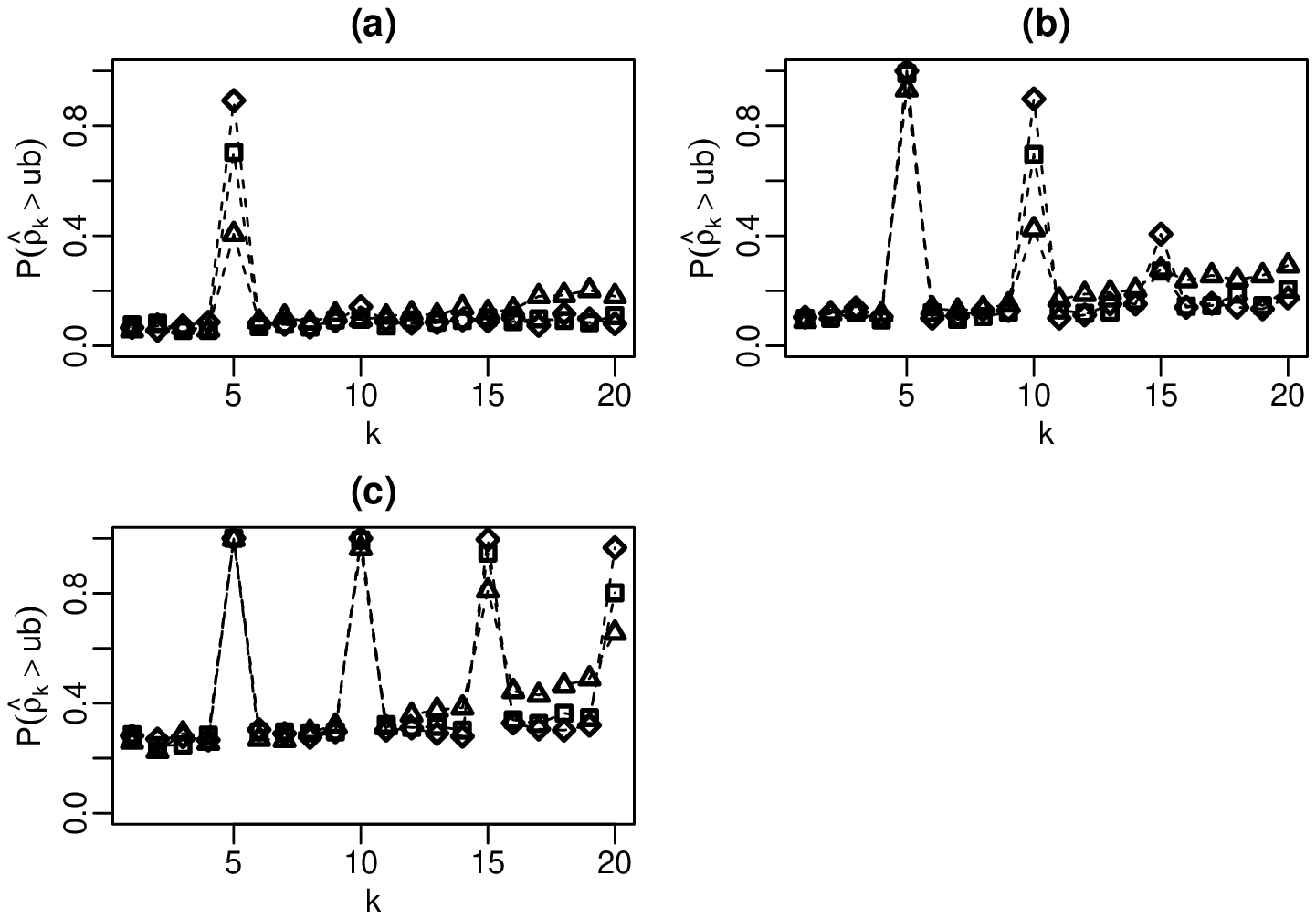';file-properties
"XNPEU";}}

\FRAME{ftbpFU}{4.7115in}{3.2716in}{0pt}{\Qcb{Simulation Results. Probability
of estimated autocorrelations $\hat{\protect\rho}_{k}$ exceeding the nominal
90\% upper confidence bound for $\mathrm{SMA}_{5}(1)$ models with parameters
(a) $b=1$ and (b) $b=3.732$, for time series lengths $n=100$ (triangles), $%
n=200$ (squares) and $n=400$ (diamonds).}}{\Qlb{fig:plots_ucb_sma}}{%
plots_ucb_sma_paper.eps}{\special{language "Scientific Word";type
"GRAPHIC";maintain-aspect-ratio TRUE;display "ICON";valid_file "F";width
4.7115in;height 3.2716in;depth 0pt;original-width 5.8894in;original-height
4.0906in;cropleft "0";croptop "1";cropright "1";cropbottom "0";filename
'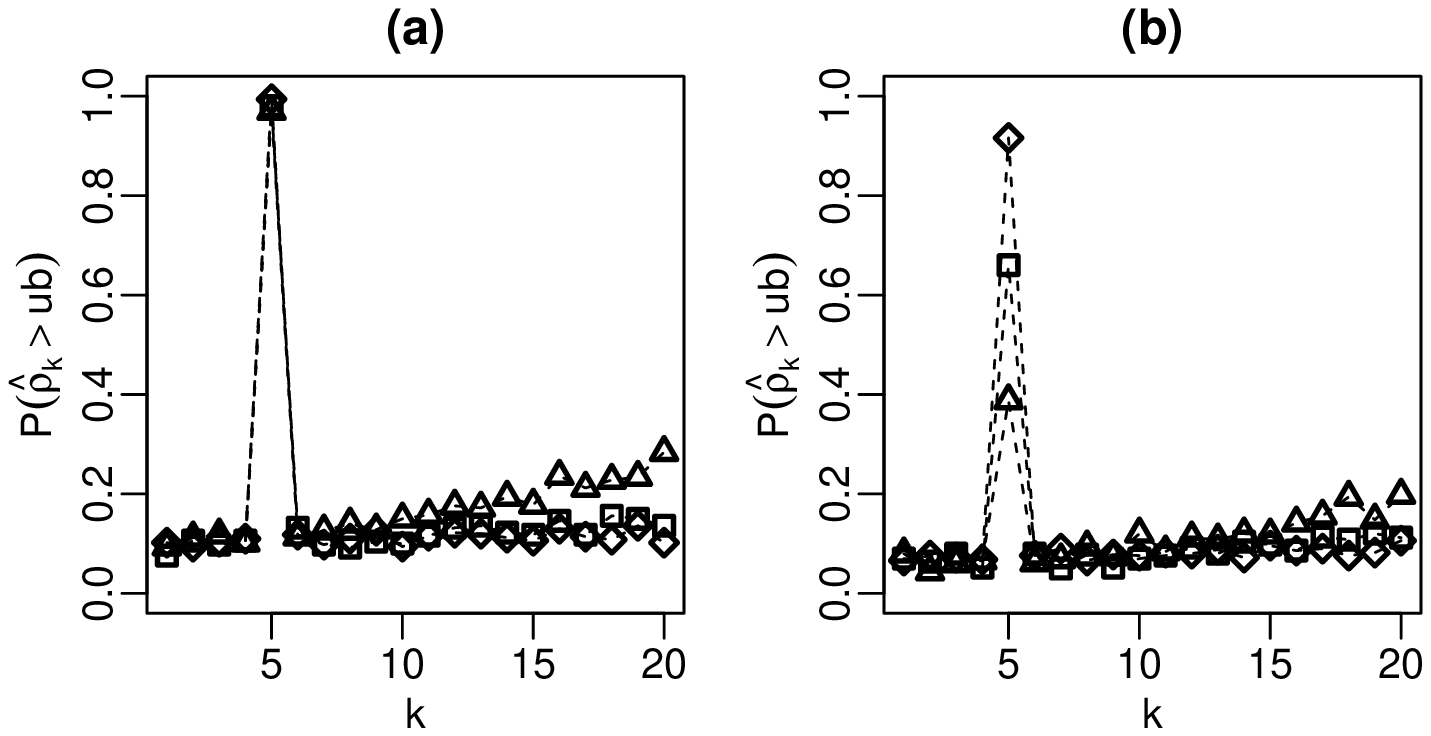';file-properties "XNPEU";}}

As before, for a few models we investigated in more depth the behavior of
the power functions $P\left( \hat{\rho}_{k}\geq \mathrm{ub}_{.10}\right) $
when both $n$ and $d$ are allowed to grow. Specifically, for the $\mathrm{AR}%
(1)$ model with $a=0.75$ and the $\mathrm{MA}\left( 1\right) $ model with $%
b=1$, we generated data with $n\in \left\{ 100,200,300,400,500\right\} $ and
computed estimators with bin sizes $d\in \left\{ 2,4,6,\ldots ,16\right\} $.
The power functions for $\hat{\rho}_{1}$, $\hat{\rho}_{2}$ and $\hat{\rho}%
_{3}$ for the $\mathrm{AR}(1)$ model, and for $\hat{\rho}_{1}$ and $\hat{\rho%
}_{2}$ for the $\mathrm{MA}\left( 1\right) $ model, are given in the
Supplementary Material. We see that the respective $\mathrm{AR}(1)$ and $%
\mathrm{MA}\left( 1\right) $ patterns will be easily detected for all $n$s
and $d$s considered, therefore it is not necessary to use bin sizes larger
than $d=5$ for the purposes of model detection.

\section{Real-data applications\label{sec:Examples}}

\subsection{Bike demand in bicycle-sharing systems}

As an example of temporal point-process time series, let us consider bicycle
check-out times in the Divvy bike-sharing system of Chicago. The data is
publicly available at the Chicago Data Portal,
https://data.cityofchicago.org. Bicycle-sharing systems have become
increasingly common in large cities around the world (Shaheen et al., 2010).
These systems provide short-term bicycle rental services at unattended
stations distributed throughout the city. For the system to run smoothly, it
is necessary that both bicycles and empty docks be available at every
station. Since bike flows from one station to another is rarely matched by a
similar flow in the reverse direction, imbalances in the spatial
distribution of bikes inevitably arise during the day (Nair and
Miller-Hooks, 2011). To manage this problem, bikes must be manually
relocated by trucks as part of the day-to-day operations of the system. From
a longer-term perspective, careful planning of the location of new stations
is important. Therefore, understanding the spatiotemporal patterns of bike
demand is fundamental for efficient planning and management of the system.

We analyzed trips that took place in 2016. For a given bike station, $X_{t}$
is defined as the set of check-out times for trips originating between 8am
and 10pm on day $t$. Of the 458 active Divvy stations in this period, we
chose two as representatives for analysis: the station at the Shedd
Aquarium, located in an area of heavy tourist traffic, and the station at
the intersection of Ashland and Wrightwood avenues, located in a typical
north-side residential neighborhood. The respective autocorrelograms, based
on five bins, are shown in Figure \ref{fig:divvy_ex}. We show
autocorrelograms for the whole annual series (Figures \ref{fig:divvy_ex}%
(a,c)) and for the three summer months from June 1 to August 31 (Figures \ref%
{fig:divvy_ex}(b,d)). The annual time series present the typical
non-stationary pattern caused by a trend. This is to be expected, since bike
trips are much more frequent in summer than in winter. Focusing on the
summer months eliminates this obvious source of nonstationarity. We can see
in Figures \ref{fig:divvy_ex}(b,d) a weekly seasonal component for both
stations, and also a small but significant spike at lag 1 for the Shedd
Aquarium station. The relatively short length of the summer series does not
allow accurate estimation at higher lags, but the non-decreasing patterns of
the spikes suggests nonstationarity of the seasonal component. Figure \ref%
{fig:divvy_ex} would then suggest the researcher $\mathrm{AR}(1)\times 
\mathrm{SAR}_{7}(1)$, $\mathrm{MA}(1)\times \mathrm{SAR}_{7}(1)$ or $\mathrm{%
SARIMA}_{7}(1,1,0)$ models for the Shedd Aquarium summer series and $\mathrm{%
SAR}_{7}(1)$ or $\mathrm{SARIMA}_{7}(1,1,0)$ models for the Ashland and
Wrightwood summer series. The annual series need to be detrended before
further modelling.

\FRAME{ftbpFU}{5.7372in}{3.7048in}{0pt}{\Qcb{Autocorrelograms for Divvy
bike-sharing data. (a) Annual time series for Shedd Aquarium station, (b)
summer time series for Shedd Aquarium station, (c) annual time series for
Ashland and Wrightwood station, and (d) summer time series for Ashland and
Wrightwood station. Dashed lines are the 5\% upper confidence bounds under
independence.}}{\Qlb{fig:divvy_ex}}{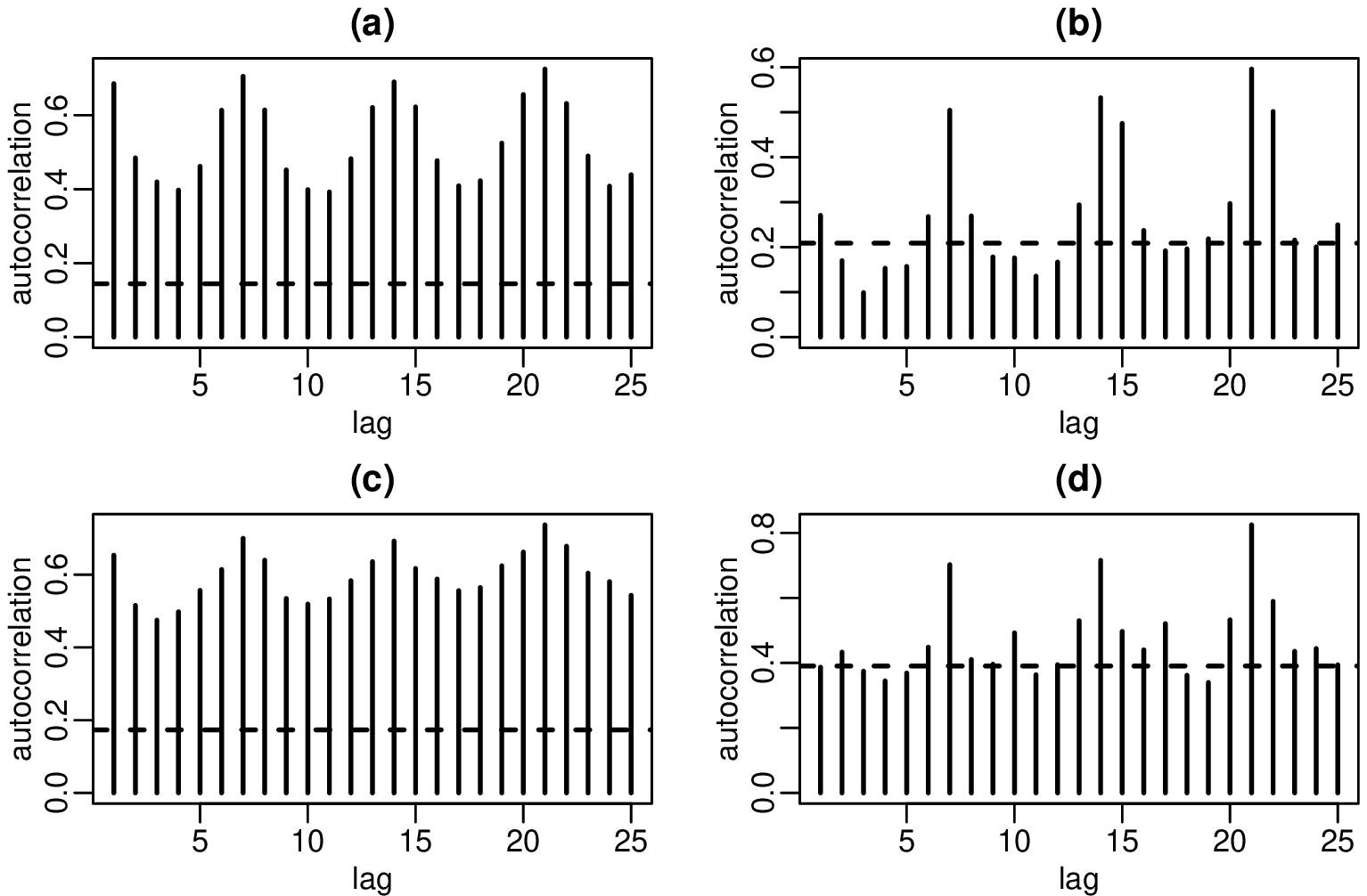}{\special{language
"Scientific Word";type "GRAPHIC";maintain-aspect-ratio TRUE;display
"ICON";valid_file "F";width 5.7372in;height 3.7048in;depth
0pt;original-width 6.3434in;original-height 4.0854in;cropleft "0";croptop
"1";cropright "1";cropbottom "0";filename
'acf_divvy_example.eps';file-properties "XNPEU";}}

These conclusions can be corroborated, to some extent, by an analysis of the
series of daily counts, which is an ordinary numerical time series. However,
the daily counts depend only on the size of the intensity functions, not
their shape, so they can only provide partial information about the process.
The autocorrelograms of log daily counts for both stations are shown in the
Supplementary Material. For the summer series of the Shedd Aquarium station
it suggests a $\mathrm{SAR}_{7}(1)$ model, in line with the above
conclusions. However, for the summer series of the Ashland and Wrightwood
station no significant autocorrelations appear at any lag. Although this
might seem to contradict Figure \ref{fig:divvy_ex}(b), the explanation is
that the weekly autocorrelations seen in Figure \ref{fig:divvy_ex}(d) are
caused by variations in the shape, not the size, of the intensity functions.
This is in line with the findings of Gervini and Khanal (2019), who showed,
using ordinary functional principal components, that the predominant mode of
variation at this station is variation in the relative proportion of morning
trips versus afternoon trips, rather than variation in the total number of
trips.

\subsection{Street theft in Chicago}

As an example of spatial time series, in this section we analyze street
robberies in Chicago during 2014. The data was also downloaded from the City
of Chicago Data Portal. There were 16,278 reported street thefts in 2014 and
their locations cover most of the city. These data can be analyzed as a
single spatiotemporal process, as Li and Guan (2014) do for similar data, or
as a time series of daily realizations of a spatial process, which is what
we do here. Then $X_{t}$ is defined as the set of spatial coordinates of the
street thefts on day $t$.

Investigating the spatial and temporal distribution of crime is important
because, as noted by Ratcliffe (2010), crime opportunities are not uniformly
distributed in space and time, and the discovery and analysis of spatial
patterns may help better understand the role of geography and opportunity in
the incidence of crime. In the specific case of street theft, it is well
known to criminologists that it tends to concentrate on places they
denominate \textquotedblleft crime attractors\textquotedblright , areas that
\textquotedblleft bring together, often in large numbers, people who carry
cash, some of whom are distracted and vulnerable\textquotedblright\
(Bernasco and Block, 2011).

\FRAME{ftbpFU}{5.3886in}{2.239in}{0pt}{\Qcb{Autocorrelograms for Chicago
street theft data. (a) Annual time series for North side, (b) annual time
series for South side. Dashed lines are the 5\% upper confidence bounds
under independence.}}{\Qlb{fig:theft_ex}}{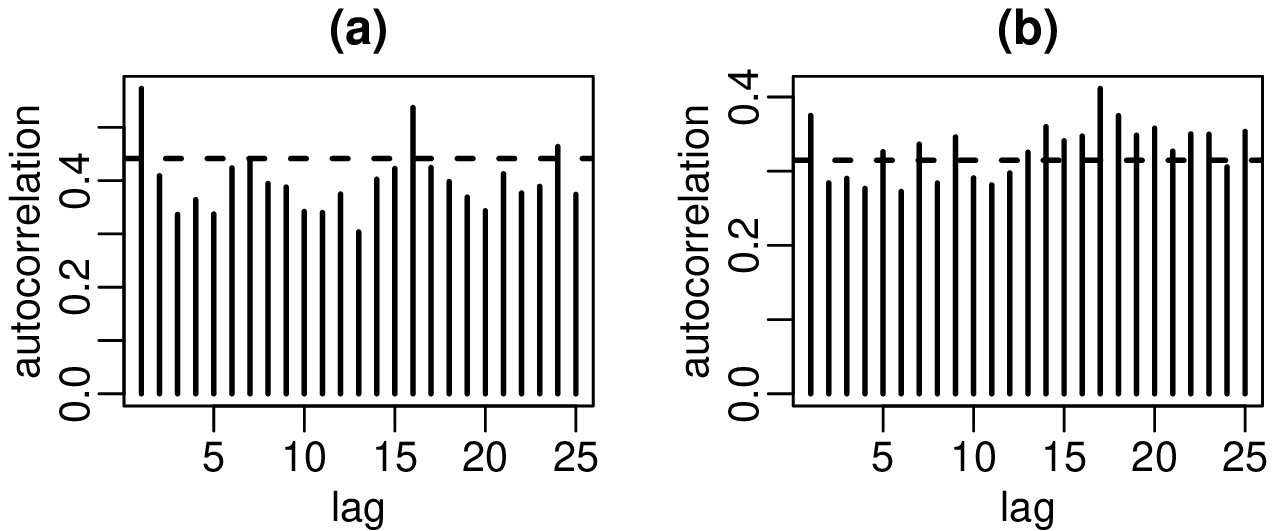}{\special%
{language "Scientific Word";type "GRAPHIC";maintain-aspect-ratio
TRUE;display "ICON";valid_file "F";width 5.3886in;height 2.239in;depth
0pt;original-width 5.361in;original-height 2.2113in;cropleft "0";croptop
"1";cropright "1";cropbottom "0";filename
'acf_theft_example.eps';file-properties "XNPEU";}}

We separately analyzed two regions of the city: the more affluent North
side, broadly defined as the area north of North Ave and East of Harlem Ave,
and the poorer South side, the area south of Roosevelt Rd and east of Cicero
Ave. The autocorrelograms, based on nine bins, are shown in Figure \ref%
{fig:theft_ex}. For the North side (Figure \ref{fig:theft_ex}(a)) we see
significant autocorrelations at lags 1 and 16; the latter is probably a
fluke, so this plot suggests an $\mathrm{MA}(1)$ model. For the South side
(Figure \ref{fig:theft_ex}(b)), on the other hand, the non-decreasing $\hat{%
\rho}_{k}$s seem to suggest nonstationarity caused by a trend. This is
confirmed by an analysis of the daily counts, shown in the Supplementary
Material: for the North side, the annual daily counts seem to be stationary
and possibly follow an $\mathrm{AR}(1)$ model, whereas for the South side
there is a clear trend that needs to be removed before further modelling.

\section{Discussion}

In this paper we presented a version of autocorrelograms for time series of
point processes. Like their univariate or functional equivalents, they are
fundamentally data-exploratory tools that point the way to more
sophisticated and rigorous data modeling. We showed by simulation that they
can successfully detect the patterns corresponding to common autoregressive
or moving average models.

The examples above also show a number of directions for further research.
First, non-stationarity due to trends was observed in both examples;
therefore, it is important to develop methods for trend estimation.
Eliminating trends by differencing, as it is commonly done for univariate,
multivariate or functional time series, is not possible for point-process
series, since the \textquotedblleft difference\textquotedblright\ between
two sets of points is not defined in any statistically meaningful way.
Second, the development of cross-correlograms would be useful for situations
where more than one process is observed at the same time; for example, for
bike-sharing systems, it may be interesting to correlate bike demand with
meteorological conditions. Finally, estimation of the autocovariance
function of the latent intensity process is an important problem in its own
right; we have avoided that problem by using a binning approach, but more
detailed information about the process can be gained by estimating the
covariance and cross-covariance functions directly, possibly by using
smoothing techniques. All this goes beyond the scope of this paper and are
topics for future research.

\section{Acknowledgement}

This work was partly supported by NSF grant DMS 2412015.

\section{References}

\begin{description}
\item Bernasco, W., and Block, R. (2011). Robberies in Chicago: a
block-level analysis of the influence of crime generators, crime attractors,
and offender anchor points. \emph{Journal of Research in Crime and
Delinquency} \textbf{48} 33--57.

\item Bouzas, P.R., Valderrama, M., Aguilera, A.M., and Ruiz-Fuentes, N.
(2006). Modelling the mean of a doubly stochastic Poisson process by
functional data analysis. \emph{Computational Statistics and Data Analysis} 
\textbf{50} 2655--2667.

\item Bouzas, P.R., Ruiz-Fuentes, N., and Oca\~{n}a, F.M. (2007). Functional
approach to the random mean of a compound Cox process. \emph{Computational
Statistics }\textbf{22} 467--479.

\item Cox, D.R., and Isham, V. (1980). \emph{Point Processes.} Chapman and
Hall/CRC, Boca Raton.

\item Diggle, P.J. (2013). \emph{Statistical Analysis of Spatial and
Spatio-Temporal Point Patterns, Third Edition.} Chapman and Hall/CRC, Boca
Raton.

\item Fern\'{a}ndez-Alcal\'{a}, R.M., Navarro-Moreno, J., and Ruiz-Molina,
J.C. (2012). On the estimation problem for the intensity of a DSMPP. \emph{%
Methodology and Computing in Applied Probability }\textbf{14} 5--16.

\item Gervini, D. (2016). Independent component models for replicated point
processes. \emph{Spatial Statistics} \textbf{18} 474--488.

\item Gervini, D. (2022a). Doubly stochastic models for spatio-temporal
covariation of replicated point processes. \emph{Canadian Journal of
Statistics} \textbf{50} 287--303.

\item Gervini, D. (2022b). Spatial kriging for replicated temporal point
processes. \emph{Spatial Statistics }\textbf{51} 100681.

\item Gervini, D. and Khanal, M. (2019). Exploring patterns of demand in
bike sharing systems via replicated point process models. \emph{Journal of
the Royal Statistical Society Series C: Applied Statistics} \textbf{68}
585--602.

\item Gervini, D. and Baur, T.J. (2020). Joint models for grid point and
response processes in longitudinal and functional data. \emph{Statistica
Sinica} \textbf{30} 1905--1924.

\item Huang, X., and Shang, H.L. (2023). Nonlinear autocorrelation function
of functional time series. \emph{Nonlinear Dynamics} \textbf{111} 2537--2554.

\item Kokoszka, P., and Reimherr, M. (2013). Determining the order of the
functional autoregressive model. \emph{Journal of Time Series Analysis} 
\textbf{34} 116--129.

\item Li, Y., and Guan, Y. (2014). Functional principal component analysis
of spatiotemporal point processes with applications in disease surveillance. 
\emph{Journal of the American Statistical Association }\textbf{109}
1205--1215.

\item Magnus, J.R., and Neudecker, H. (1999). \emph{Matrix Differential
Calculus with Applications in Statistics and Econometrics. Revised Edition.}
Wiley, New York.

\item Mestre, G., Portela, J., Rice, G., Roque, A.M.S., Alonso, E. (2021).
Functional time series model identification and diagnosis by means of auto-
and partial autocorrelation analysis. \emph{Computational Statistics and
Data Analysis} \textbf{155} 107108.

\item M\o ller, J., and Waagepetersen, R.P. (2004). \emph{Statistical
Inference and Simulation for Spatial Point Processes}. Chapman and Hall/CRC,
Boca Raton.

\item Nair, R., and Miller-Hooks, E. (2011). Fleet management for vehicle
sharing operations. \emph{Transportation Science }\textbf{45 }524--540.

\item Ratcliffe, J. (2010). Crime mapping: spatial and temporal challenges.
In \emph{Handbook of Quantitative Criminology}, A.R. Piquero and D. Weisburd
(eds.), pp.~5--24. New York: Springer.

\item Shaheen, S., Guzman, S., and Zhang, H. (2010). Bike sharing in Europe,
the Americas and Asia: Past, present and future. \emph{Transportation
Research Record: Journal of the Transportation Research Board }\textbf{2143}
159--167.

\item Snyder, D.L., and Miller, M.I. (1991). \emph{Random Point Processes in
Time and Space.} Springer, New York.

\item Streit, R.L. (2010). \emph{Poisson Point Processes: Imaging, Tracking,
and Sensing.} Springer, New York.

\item Wu, S., M\"{u}ller, H.-G., and Zhang, Z. (2013). Functional data
analysis for point processes with rare events. \emph{Statistica Sinica }%
\textbf{23} 1--23.
\end{description}

\end{document}